\documentclass[10pt,journal,draftcls,letterpaper,onecolumn]{IEEEtran}
\usepackage{amssymb}
\usepackage{amsmath}
\usepackage{epsfig}
\usepackage{algorithm}
\usepackage{graphicx}
\usepackage{subfigure}
\usepackage{bm}
\usepackage{amsthm}
\usepackage{multirow}
\usepackage{tikz}
\usetikzlibrary{arrows.meta}
\usetikzlibrary{bending}
\usetikzlibrary{topaths}

\newtheorem{myProp}{Proposition}
\newtheorem{myProf}{Proof}

\allowdisplaybreaks

\begin{document}

\title{Underlaid Sensing Pilot for Integrated Sensing and Communications}
\author{\IEEEauthorblockN{Pu~Yuan\IEEEauthorrefmark{1},
Hao~Liu\IEEEauthorrefmark{1}, Junjie~Tan\IEEEauthorrefmark{1}, Dajie~Jiang\IEEEauthorrefmark{1}, Lei~Yan\IEEEauthorrefmark{2}\\}
\IEEEauthorblockA{\IEEEauthorrefmark{1}vivo Mobile Communications Co.,Ltd., Beijing, China 100015\\} \IEEEauthorblockA{\IEEEauthorrefmark{2}Northeastern University, Qinhuangdao, China 066000\\}
%\author{\IEEEauthorblockN{Pu~Yuan, Hao~Liu, Junjie~Tan, Dajie~Jiang\\}
%\IEEEauthorblockA{vivo Mobile Communications Co.,Ltd., Beijing, China 100015\\}
%Email: \{yuanpu, hao.liu, junjie.tan jiangdajie\}@vivo.com}
%\author{\IEEEauthorblockN{Pu~Yuan, Author1, Author2, Author3\\}
%\IEEEauthorblockA{vivo Mobile Communications Co.,Ltd., Beijing, China 100015\\}
Email: \{yuanpu, hao.liu, junjie.tan jiangdajie\}@vivo.com, yanlei@qhd.neu.edu.cn}
\maketitle

\begin{abstract}
This paper investigates a novel underlaid sensing pilot signal design for integrated sensing and communications (ISAC) in an OFDM-based communication system. The proposed two-dimensional (2D) pilot signal is first generated on the delay-Doppler (DD) plane and then converted to the time-frequency (TF) plane for multiplexing with the OFDM data symbols. The sensing signal underlays the OFDM data, allowing for the sharing of time-frequency resources. In this framework, sensing detection is implemented based on a simple 2D correlation, taking advantage of the favorable auto-correlation properties of the sensing pilot. In the communication part, the sensing pilot, served as a known signal, can be utilized for channel estimation and equalization to ensure optimal symbol detection performance. The underlaid sensing pilot demonstrates good scalability and can adapt to different delay and Doppler resolution requirements without violating the OFDM frame structure. Experimental results show the effective sensing performance of the proposed pilot, with only a small fraction of power shared from the OFDM data, while maintaining satisfactory symbol detection performance in communication.

\end{abstract}

%Conventional approach usually adopts a high power pulse as the pilot while simple power detector can be utilized for channel estimation. However, this design results in an imbalance power distribution among different delay-taps in the delay-Doppler plane, and consequently the periodical high power samples in the time domain waveform.

\section{Introduction}\label{sec:sec0}
The concurrent ISAC systems can be categorized into two main types: data-based sensing and pilot-based sensing. The latter exhibits better resilience to interferences in multi-user systems, while the commonly known sensing pilot enables multiple-station sensing. Pilot-based ISAC consists of two major components: the sensing signal and the data signal, usually multiplexed using frequency division multiplexing (FDM) or time division multiplexing (TDM) in conventional ISAC systems. In this paper, we propose a novel code division multiplexing (CDM)-like ISAC system in which the sensing pilot and data share the same time-frequency resources but are modulated and detected in different domains. The key features of the proposed scheme are as follows:
\begin{itemize}
\item[-] \textbf{\textit{The scalability.}} 
The underlaid sensing pilot can scale across multiple OFDM slots and sub-bands without violating the numerology and resource allocation of the overlaid communication system. It can adapt to various sensing requirements, and the sensing detection complexity scales linearly with the pilot size.
\item[-] \textbf{\textit{The flexibility.} }
The underlaid 2D pilot can be sparse in the delay-Doppler plane for specific purposes such as energy saving or interference avoidance. Additionally, multiple antenna ports are supported, allowing the 2D pilots to be multi-layered, with each layer corresponding to an antenna to enable angle of arrival (AoA)/angle of departure (AoD) estimation.
\item[-] \textbf{\textit{The quasi-orthogonality.}}
The underlaid sensing pilot can be generated from a series of sequences to maintain a low cross-correlation with the data signal. It can be viewed as a code division multiplexing technique where the power of the sensing and communication signals is projected into different subspaces via different codewords, thus avoiding mutual interference.
\item[-] \textbf{\textit{The separability.}}
Interference caused by the sensing pilot on the data can be mitigated with the provision of side information. Firstly, the known time-frequency domain sensing pilot can act as part of the reference signal (RS) for channel estimation, thereby maintaining the signal-to-interference-plus-noise ratio (SINR) of the RS. Secondly, upon obtaining the channel state information (CSI), the communication receiver can easily remove the interference contributed by the sensing pilot.
\end{itemize}

\section{ISAC Signal Transmitter Design}\label{sec:sec1}
The design philosophy of the 2D pilot signal follows the fact that the channel coupling effect on the transmitted signal occurs in the delay-Doppler domain through a twisted convolution mechanism \cite{mohammed2022otfs}. Specifically, coupled with the channel, the received pilot signal exhibits cyclic shifts in both the Doppler and delay dimensions, as well as phase offsets due to baseband processing of the transmit pulse\cite{tse2005fundamentals}. Moreover, conventional pilot designs in the delay-Doppler domain can lead to the peak-to-average power ratio (PAPR) problem\cite{yuan2022lpp}. To address this issue, we propose a 2D pilot design that spreads the pilot power throughout the delay-Doppler plane, resulting in reduced magnitude variation of the time domain samples.

\subsection{Sensing Pilot Design and Property}
To leverage the twisted convolution of DD domain channel in sensing detection, the following properties of the 2D pilot are demand. Firstly, the 2D pilot should have good auto-correlation to its cyclic-shifts in both dimensions. Secondly, the 2D pilot should have good cross-correlation to the data signal. Thirdly, the 2D pilot should have acceptable PAPR and ease for detection. 

Without loss of generality, suppose we have two sequences $\mathbf{a}=[a_1,a_2,...,a_Q]$ and $\mathbf{b}=[b_1,a_2,...,b_P]^H$ who are with good auto and cross correlation. $\forall 0\leq p,i<P, 0\leq q,j<Q, p,i,q,j\in \mathbb{N}$, the following equations hold.
\begin{equation}
\textrm{tr}\left(\frac{1}{Q}\mathbf{a}^H_{[q]}\mathbf{a}_{[j]}\right)
 = \frac{1}{Q}\mathbf{a}_{[q]}\mathbf{a}_{[j]}^H = 
\left\{
\begin{array}{cc}
1, & q=j,  \\
\epsilon_a & q\neq j.   \\
\end{array}\right.
\label{eq:eq0}
\end{equation}

\begin{equation}
\textrm{tr}\left(\frac{1}{P}\mathbf{b}_{[p]}\mathbf{b}^H_{[i]}\right)
 = \frac{1}{P}\mathbf{b}^H_{[p]}\mathbf{b}_{[i]} = 
\left\{
\begin{array}{cc}
1, & p=i,  \\
\epsilon_b & p\neq i.  \\
\end{array}\right.
\label{eq:eq1}
\end{equation}
where $(\cdot)^H$ denots the transpose, $(\cdot)_{[i]}$ denotes the cyclic-shift of a vector and $0\leq \left|\epsilon_a\right|, \left|\epsilon_q\right| \ll 1$. 

The 2D pilot can be generated from the Kronecker product or product of the $\mathbf{a}$ and $\mathbf{b}$.
\begin{equation}
\mathbf{P} = \mathbf{a}\otimes \mathbf{b} = \mathbf{b}\mathbf{a}.
\label{eq:eq2}
\end{equation}
Consequently we define the 2D cyclic shift $\mathbf{P}_{[q,p]}$ as the follows,
\begin{equation}
\mathbf{P}_{[q,p]} = \mathbf{a}_{[q]}\otimes \mathbf{b}_{[p]} = \mathbf{b}_{[p]}\mathbf{a}_{[q]} .
\label{eq:eq3}
\end{equation}
where $(\cdot)_{[i,j]}$ denotes cyclic shifting $i$ and $j$ units in row and column. Based on above notations, the correlation matrix of $\mathbf{P}$ and $\mathbf{P}_{[q,p]}$ has the property depict in proposition \ref{prop:prop1}.
\begin{myProp}
The inner product of $\mathbf{P}_{[q,p]}$ and $\mathbf{P}_{[j,i]}$ yields the following values,
\begin{equation}
\frac{1}{QP}\left<\mathbf{P}_{[q,p]},\mathbf{P}_{[j,i]}\right> = 
%\textrm{tr}(\frac{1}{QP}\mathbf{P}_{[q,p]}\mathbf{P}^H_{[j,i]}) = 
\left\{
\begin{array}{cc}
1, & q=j, p=i  \\
\epsilon_p & q=j, p\neq i   \\
\epsilon_q & q\neq j, p=i   \\
\epsilon_q\epsilon_p & q\neq j, p\neq i   \\
\end{array}\right.
\label{eq:eq4}
\end{equation}\label{prop:prop1}
\end{myProp}

Proposition \ref{prop:prop1} is proved as follows, 
\begin{myProf}
The inner product of two matrices has the following property,
\begin{align}
\left<\mathbf{A},\mathbf{B}\right>
 = \mathrm{tr}\left(\mathbf{A}^H\mathbf{B}\right)
\label{eq:eq5}
\end{align}
therefore we have the follows,
\begin{align}
&\frac{1}{QP}\left<\mathbf{P}_{[q,p]},\mathbf{P}_{[j,i]}\right>
 = \mathrm{tr}\left(\frac{1}{QP}\mathbf{P}^H_{[q,p]}\mathbf{P}_{[j,i]}\right) \nonumber\\ 
& = \mathrm{tr}\left(\frac{1}{QP}\mathbf{a}^H_{[q]}\mathbf{b}^H_{[p]}\mathbf{b}_{[j]}\mathbf{a}_{[i]}\right) 
 = \mathrm{tr}\left(\frac{1}{Q}\mathbf{b}^H_{[p]}\mathbf{b}_{[j]}\frac{1}{P}\mathbf{a}^H_{[q]}\mathbf{a}_{[i]}\right) \nonumber\\ 
& = \frac{1}{P}\mathbf{b}^H_{[p]}\mathbf{b}_{[j]}\mathrm{tr}\left(\frac{1}{Q}\mathbf{a}^H_{[q]}\mathbf{a}_{[j]}\right).
\label{eq:eq6}
\end{align}
Substitute equation (\ref{eq:eq0}) and (\ref{eq:eq1}) into (\ref{eq:eq6}), we obtain the results in proposition \ref{prop:prop1}.
\qed
\end{myProf}

Proposition \ref{prop:prop1} implies an efficient 2D correlation based detection at the sensing receiver, which will be discussed in section \ref{sec:sec2}. 

\subsection{ISAC Signal Generation}\label{sec:sec2}
\begin{figure}
\centering
\includegraphics[width=0.4\textwidth]{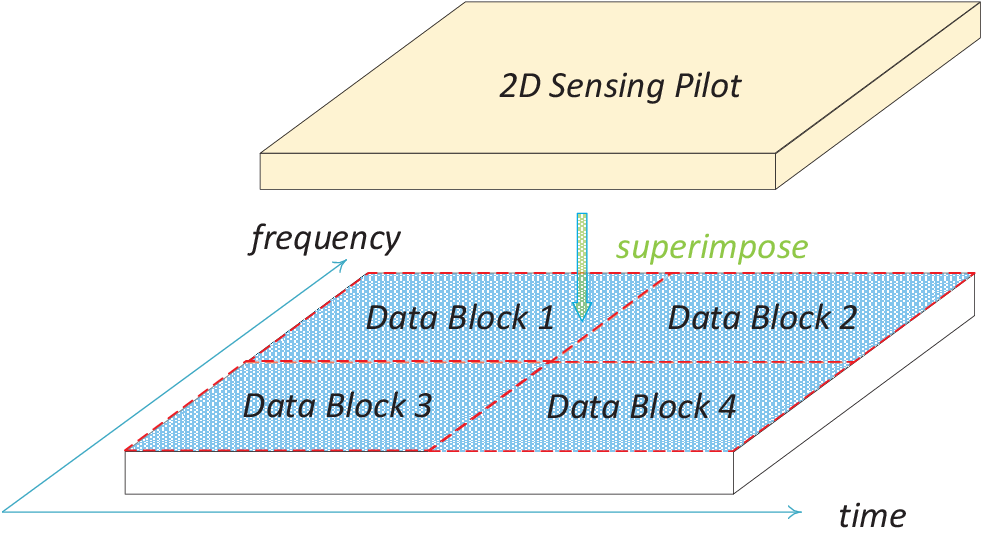}
  \caption[]{The underlaid sensing pilot with OFDM data.}
	\label{fig:fig5}
\end{figure}

We consider a frame size of $M\times N$, where $N$ and $M$ corresponds to the number of Doppler and delay taps in delay-Doppler plane, as well as the number of OFDM symbols and sub-carriers in the time frequency plane, respectively. The Doppler and delay are quantized as $\tau=\frac{T}{M}$ and $\nu=\frac{1}{N\Delta f}$. We denote the symbol time and subcarrier spacing of the OFDM as $T$ and $\Delta f$. We denote the discrete samples of DD domain sensing pilot as $\mathbf{P}\in\mathbb{C}^{M \times N}$, and the discrete samples of TF domain data as $\mathbf{D}\in\mathbb{C}^{M \times N}$. The proposed sensing pilot design can be directly incorporate into the OFDM system without violating the existing design. The 2D pilot will firstly be transformed from DD plane to TF plane via ISFFT, then superimposed with the OFDM data symbols in the TF plane as illustrated in figure \ref{fig:fig5}. Here the discrete TF domain form of the superimposed signal is given by,
\begin{equation}
\mathbf{X} = \mathbf{F}_M\mathbf{P}\mathbf{F}^H_N + \mathbf{D}.
\label{eq:eq7}
\end{equation}
, where $\mathbf{F}_M$ denotes the M-point DFT matrix. 

Note that the RS for channel estimation is also part of the data matrix $\mathbf{D}$. The superimposed TF domain samples are converted to the time-delay plane and appended with symbols-wise cyclic prefix (CP) as follows,
\begin{equation}
\tilde{\mathbf{X}} = \mathbf{B}_{cp}\mathbf{F}^H_M\mathbf{X} = \mathbf{B}_{cp}\left(\mathbf{P}\mathbf{F}^H_N + \mathbf{F}^H_M\mathbf{D}\right),
\label{eq:eq8}
\end{equation}
where $\mathbf{B}_{cp} = \left[
\begin{array}{ccc}
\mathbf{0}_{L_{cp}\times (M-L_{cp})} & \mathbf{I}_{L_{cp}}  \\
  \mathbf{I}_{M} &   \\
\end{array}
\right]$
is the operator for appending a CP of $L_{cp}$ samples \cite{9303350}. 
Then the discrete time signal $\tilde{\mathbf{x}}$ is give by vectorizing the time-delay plane samples and impose with a prototype filter $\mathbf{G}_{tx}$ , 
\begin{equation}
\tilde{\mathbf{x}} = \left(\mathbf{F}^H_N\otimes\mathbf{G}_{tx}\right)\mathbf{vec}\left(\mathbf{B}_{cp}\mathbf{P}\right) + \mathbf{vec}\left(\mathbf{G}_{tx}\mathbf{B}_{cp}\mathbf{F}^H_M\mathbf{D}\right).
\label{eq:eq9}
\end{equation}
If following the conventional OFDM operation where rectangular pulse is applied, then we have $\mathbf{G}_{tx}=\mathbf{I}_M$. 
In practical system, $\tilde{\mathbf{x}}$ is up-sampled in the DAC to generate the continuous time signal. Without loss of generality, we limited our analysis with the discrete time model and abbreviate the discussion on exact time domain waveform or the prototype filter supposing perfect synchronization and best sampling.

\section{Sensing and Communication Receiver Processing}\label{sec:sec3}
\subsection{Sensing Pilot Detection}
At the sensing receiver, after ADC and down sampling, the DD domain received pilot $\mathbf{R}$ is actually an cyclic-shifted, phase-offset and power attenuated version of the original one. Denote $r[k,l]$ and $p[k,l]$ as the elements of $\mathbf{R}$ and $\mathbf{P}$ respectively, we can write the DD domain input-output relationship following the same line as \cite{Raviteja2018interference},
\begin{equation}
r[k,l] = \sum^{L}_{i=1}{h_i\alpha_{i}(k,l)(p[[k-k_i]_N,[l-l_i]_M])e^{j2\pi\frac{(L_{cp}+l-l_i)k_i}{N(M+L_{cp})}}}, 
\label{eq:eq10}
\end{equation}
where $h_i$ is the channel attenuation factor of the path $i$ and 
$\alpha_{i}(k,l)  = \left\{
\begin{array}{cc}
1 & l_i\leq l<M  \\
\frac{N-1}{N}e^{-j2\pi\frac{[k-k_i]_N}{N}} &  0\leq l<l_i
\end{array}\right.
$
is the phase offset term due to the rectangular pulse shaping of OFDM, and the impact of CP on the phase term is derived from the result in \cite{9303350}. Note that (\ref{eq:eq10}) neglects the impact of fractional Doppler by assuming large $M$ and $N$ such that the resolution of Doppler and delay are fairly enough.

In sensing detection we usually assume that each echo path corresponds to one reflector, which can be modeled as,
\begin{equation}
\tilde{p}[k,l] = h_i\beta_{i}(k,l)(p[k,l]\ast \delta[[k-k_i]_N,[l-l_i]_M]), 
\label{eq:eq11}
\end{equation}
where $\beta_{i}(k,l) = \alpha_{i}(k,l)e^{j2\pi\frac{(L_{cp}+l-l_i)k_i}{N(M+L_cp)}}$ is the time varying phase change caused by the Doppler, $p[k,l]\ast \delta[[k-k_i]_N,[l-l_i]_M]$ is the cyclic-shift operation corresponds to $\mathbf{P}_{[k,l]}$.

Since the attenuation factor $h_i$ is irrelevant to Doppler and delay detection, we can approximately give the equivalent matrix form input-output of each echo path by abbreviating $h_i$,
\begin{equation}
\mathbf{R} = \mathbf{P}_{[q,p]}\odot \Xi_{<q,p>} , 
\label{eq:eq12}
\end{equation}
where $\odot$ denotes the Hadamard product, i.e., the point-wise product.
In (\ref{eq:eq11}), $\Xi_{<q,p>}$ is the phase offset matrix applied to DD domain discrete samples with the $\beta_{i}(k,l)$ as its entries. 

Note that the inner product between $\mathbf{P}_{[q,p]}$ and $\mathbf{P}_{[j,i]}$ gives the maximum when $j=q, i=p$, 
\begin{align}
& \mathrm{arg max} \left<\mathbf{P}_{[q,p]},\mathbf{P}_{[i,j]}\right> \nonumber \\ 
& = \sum^N_{n=1}{\sum^M_{m=1}{\left[\mathbf{P}_{[q,p]}\odot \mathbf{P}_{[i,j]}\right]_{(n,m)}}} |_{q=j,p=i}\nonumber \\ 
& = \sum^N_{n=1}{\sum^M_{m=1}{\left[\mathbf{R}\odot \Xi^{*}_{<q,p>}\odot \mathbf{P}_{<j,i>}\right]_{(n,m)}}}|_{q=j,p=i} \nonumber \\ 
& = \sum^N_{n=1}{\sum^M_{m=1}{\left[\mathbf{R}\odot \Xi^{*}_{<j,i>}\odot \mathbf{P}_{<j,i>}\right]_{(n,m)}}}|_{q=j,p=i}.
\label{eq:eq13}
\end{align}

Equation (\ref{eq:eq13}) explicitly gives the structure of the proposed sensing detector, which can be summarized as a series of hypothesis test of the received DD domain discrete samples with the cyclic shifted and phase compensated version of the original 2D pilot.

In algorithm $\mathbf{1}$, the receiver will traverse over all possible pairs of $(q,p)$ to calculate the inner product of the corresponding $\Xi^{*}_{<q,p> \odot \mathbf{P}}$ with $\mathbf{R}$, and take the one pair who leads to the locally maximum as the sensing result. Therefore, the sensing detector is fairly simple since it requires only shift and product-sum operations.

\begin{algorithm}
\textbf{Initialization} : $k_{max}$, $l_{max}$, $\mathbf{P}$, $\gamma$.

\textbf{Detection} :
\begin{itemize}
\item[a)] $\forall k\leq k_{max}, l\leq l_{max}$, calculate the detection matrix $\mathbf{P}_{det}$ for each Doppler and delay pair $<k, l>$
\begin{equation}
\mathbf{P}_{det} = \mathbf{P}_{[k,l]}\odot \Xi_{<k,l>}, \nonumber
\label{eq:pd1}
\end{equation}

\item[b)] Calculate the absolute value of matrix correlation,
\begin{align}
vd_{<k,l>} = \left|\sum^N_{n=1}{\sum^M_{m=1}{\left[\mathbf{R}\odot \Xi^{*}_{<k,l>}\odot \mathbf{P}_{det}\right]_{(n,m)}}}\right|. \nonumber
\label{eq:pd2}
\end{align}

\item[c)] 
$\forall k, l$, where $vd_{<k,l>} >\gamma$, then the echo paths with Doppler $k$ and delay $l$ are identified, i.e., a set of reflectors with velocity and distance $<k\frac{1}{NT},l\frac{1}{M\Delta f}>$ is detected.
\end{itemize}
\label{alg:alg1}
\caption{Sensing Detection}
\end{algorithm}

Note that the phase compensation before the 2D correlation is optional if phase noise resilient sequences are utilized in the 2D pilot. In such case the phase noise only produces fairly small sidelobe adjacent to the correlation peak, and the negative impact on distance and velocity detection can be negligible. It is different from the channel estimation in communication where the phase information is demanded for the symbol demodulation.

According to equation (\ref{eq:eq5}), the correlation peak only occurs when the two matrices cyclic shift the same in both dimensions. Otherwise, the value of the correlator output is fairly small. This fact guarantees the performance of the threshold based decision making, where no correlation peak will be pass the threshold if the under tested Doppler and delay pair $<k,l>$ does not correctly match with the $(\nu, \tau)$ of the echo path. In case of fractional Doppler and delay, the detection results $<k,l>$ could be further refined to reduce the estimation error, which will be discussed in section \ref{sec:sec4}.

\subsection{Communication Receiver}
The received signal after ADC and down sampling yields the discrete time samples, where the equivalent time domain input-output in (\ref{eq:eq8}) of each OFDM symbol can be written in vector form as $\mathbf{r} = \mathbf{H_t}\mathbf{s}+\mathbf{w}$ where $\mathbf{H_t} = \sum^L_{i=1}{h_i\bm{\Pi}^{l_i}\bm{\Delta}^{(k_i)}}, \mathbf{H_t}
 \in \mathbb{C}^{M\times M}$ is the equivalent time domain channel matrix derived from path delay, Doppler, channel gain and shaping pulse in the same way as \cite{Raviteja2018interference}. Hence the received samples per OFDM symbol after removing CP is given by,
\begin{equation}
\mathbf{r} = \mathbf{H_t}\left(\mathbf{p}\mathbf{F}^H_N+\mathbf{F}^H_M\mathbf{d}\right)+\mathbf{w},
\label{eq:eq14}
\end{equation}
where $\mathbf{p}$ and $\mathbf{d}$ are the columns of $\mathbf{P}$ and $\mathbf{D}$ respectively.

In OFDM system we usually impose a comb-like RS for channel estimation, which is a pre-defined sequence known for both transmitter and receiver. The elements corresponds to the position of the RS in the TF domain channel matrix is precisely estimated, while others are estimated using some interpolation schemes. Upon obtaining the CSI, a TF domain single-tap equalization is followed to decouple the channel. If the MMSE criterion is adopted, the OFDM detection can be written as,
\begin{equation}
\hat{\mathbf{x}} = (\mathbf{h}^{*}_{tf}\odot (\mathbf{F}_{M}\mathbf{r}))\oslash (\left|\mathbf{h}_{tf}\right|^{2}+\sigma^2),
\label{eq:eq15}
\end{equation}
where $\sigma^2$ is the noise variance, $\mathbf{h}_{tf}=\mathrm{diag}\left(F^H_M\mathbf{H_t}F_M\right)$ is the TF domain channel, $\oslash$ denotes the Hadamard division.

Essentially, the performance of data symbol detection in (\ref{eq:eq15}) relies on the estimated channel $\mathbf{h}_{tf}$. If the protocol enables the knowledge on sensing pilot at the communication receiver, the underlaid sensing pilot would not violate the channel estimation accuracy. 

Algorithm $2$ suggests a possible way of utilizing the sensing pilot at the communication receiver. In a nutshell, we can use the equivalent RS which is the summation of the original RS and co-located sensing pilot elements to guarantee an unbiased estimate of $\mathbf{h}_{tf}$, followed by a sensing pilot cancellation to ensure a low residual interference during symbol detection. 

In algorithm $2$ we denote $\mathbf{Y}$ as the TF domain discrete samples of the received signal, $\mathrm{supp}(\mathbf{A})$ as the support of matrix $\mathbf{A}$, which is the subset of $\mathbf{A}$ with all zero entries removed. 
We denote $\mathbf{T}$ is the $M \times N$ index matrix of the OFDM RS, where the $1$ entries indicates the position of RS and $0$ for the data, hence the RS set in the TF plane can be denoted by $\mathbf{D}\odot \mathbf{T}$.
\begin{algorithm}
\textbf{Initialization} : $\mathbf{T}$, $\mathbf{D}$, $\mathbf{P}$.

\textbf{Channel Estimation} :
\begin{itemize}
\item[a)] Equivalent RS $\mathbf{S}_{eff}$ is calculated by,
\begin{equation}
\mathbf{S}_{eff} = \mathrm{supp}\left(\mathbf{D}+\mathbf{F}_M\mathbf{P}\mathbf{F}^H_N\odot \mathbf{T}\right), \nonumber
\label{eq:pd3}
\end{equation}
\item[b)] The least square (LS) channel estimator is give by,
\begin{align}
\hat{\mathbf{H}}_{eff} = \mathrm{supp}\left(\mathbf{Y}\odot \mathbf{T} \right)\oslash \mathbf{S}_{eff}. \nonumber
\label{eq:pd4}
\end{align}
Then $\hat{\mathbf{H}}_{eff}$ can be further refined (e.g. using MMSE) and interpolated to channel matrix $\hat{\mathbf{H}}_{tf}\in \mathbb{C}^{M\times N}$. 
\end{itemize}
\textbf{Symbol Detection} :
\begin{itemize}
\item[a)] 
The sensing signal is canceled as a interference followed by one-tap equalization.
\begin{align}
\hat{\mathbf{D}} = \hat{\mathbf{H}}_{tf}^{*}\odot\left(\mathbf{Y}-\mathbf{F}_M\mathbf{P}\mathbf{F}^H_N \odot \hat{\mathbf{H}}_{tf} \right)
\oslash \left(\left|\hat{\mathbf{H}}_{tf}\right|^2+\mathbf{\sigma}^2 \right). \nonumber
\end{align}
\end{itemize}
\label{alg:alg2}
\caption{Symbol Detection}
\end{algorithm}

\section{Sensing Performance Analysis}\label{sec:sec4}
\subsection{Sensing SINR}
For sensing detection, we usually considers the LOS path to the sensing target, while the NLOS path or path to other reflectors as the interfering paths. Hence the received signal can be decomposed into three components,
\begin{align}
%\begin{split}
&\mathbf{R} = \sum^{L-1}_{i=0}{h_i(\mathbf{P}_{[l_i,k_i]}+(\mathbf{F}^H_M\mathbf{P}\mathbf{F}_N)_{[l_i,k_i]})} + \mathbf{W} \\ \nonumber
&= \underbrace{h_0\mathbf{P}_{[l_0,k_0]}}_{Signal} 
+ \underbrace{\sum^{L-1}_{i=1}{h_i\mathbf{P}_{[l_i,k_i]}}+\sum^{L-1}_{i=0}{h_i(\mathbf{F}^H_M\mathbf{D}\mathbf{F}_N)_{[l_i,k_i]}} }_{Interference}
+ \underbrace{\mathbf{W}}_{Noise}.
%\end{split}
\label{eq:eq16}
\end{align}

As described in section \ref{sec:sec2}, the sensing detection is implemented by feeding the received DD domain signal $\mathbf{R}$ and $\mathbf{P}_{[l,k]}$ into a 2D correlator, and the correlation peak occurs when the locally generated matrix $\mathbf{P}_{[l_0,k_0]}$ is chosen as an input. In such case, the output is given by, 
\begin{align}
%\begin{split}
& <\mathbf{P}_{[l_0,k_0]},\mathbf{Y}> \triangleq MNh_0+\sum^{L-1}_{l=1}{h_i<\mathbf{P}_{[l_0,k_0]},\mathbf{P}_{l_i,k_i}}> \\ \nonumber
& +\sum^{L-1}_{l=0}{h_i<\mathbf{P}_{[l_0,k_0]},\mathbf{D}_{l_i,k_i}}>+<\mathbf{P}_{[l_0,k_0]},\mathbf{W}^H>.
%\end{split}
\label{eq:eq17}
\end{align}

It is noted that in (\ref{eq:eq17}) the desired output value $MNh_0$ is distorted by the interference and noise terms, therefore false detection may occurs in case of serious distortion. 
Denote $\vartheta_{0i}$, $\rho_{0i}$ and $\varsigma_{0i}$ as 
$<\mathbf{P}_{[l_0,k_0]},\mathbf{P}_{l_i,k_i}>$, 
$<\mathbf{P}_{[l_0,k_0]},\mathbf{D}_{l_i,k_i}>$ and 
$<\mathbf{P}_{[l_0,k_0]},\mathbf{W}^H>$ respectively, we define the sensing SINR $Z$ of the proposed scheme as follows, 
\begin{equation}
%\begin{split}
Z \triangleq \frac{MNh_0}{\sum^{L-1}_{l=1}{h_i\vartheta_{0i}} +\sum^{L-1}_{l=0}{h_i\rho_{0i}}+\varsigma_{0i}},
%\end{split}
\label{eq:eq18}
\end{equation}
The sensing SINR here actually characterizes the distortion level, and the performance is affected by the noise and interference via $\rho_{0i}$ and $\varsigma_{0i}$. Note that $\vartheta_{0i}$ can be neglect since the auto correlation of pilot with its cyclic-shifts is insignificant. For $\rho_{0i}$ and $\varsigma_{0i}$, we experimentally show they are also insignificant by providing the power distribution of the the two variables.

\begin{figure}
\centering
\subfigure[M=15, N=15]{\includegraphics[width=0.45\textwidth]{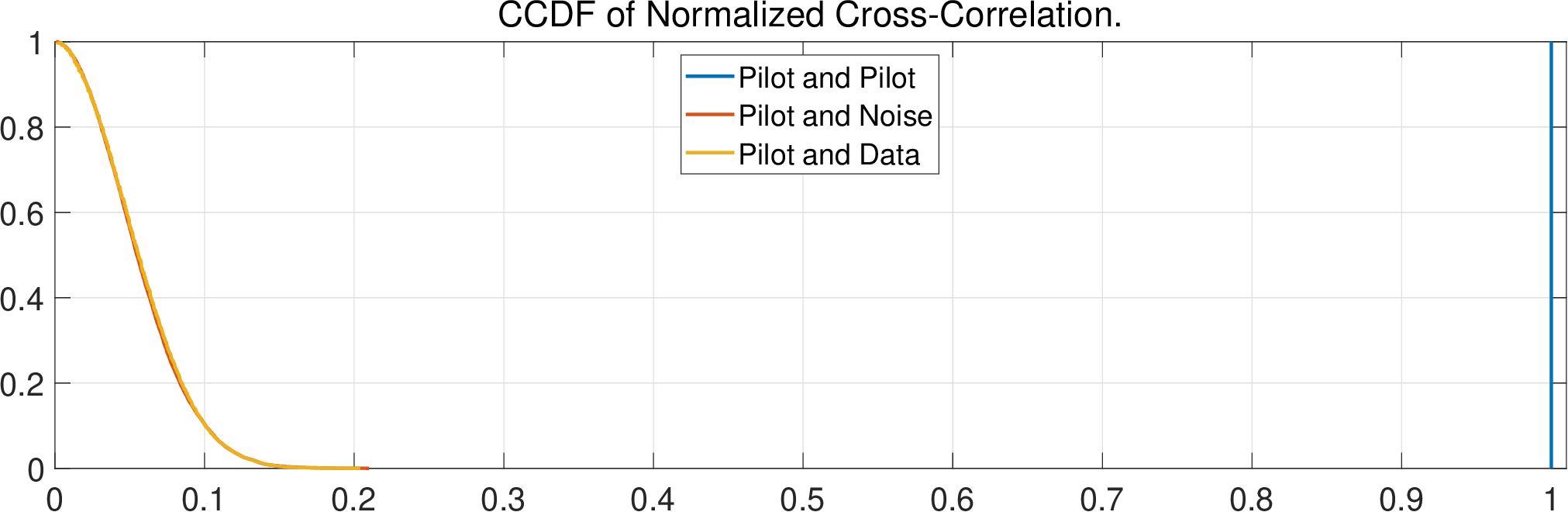}}\\
\subfigure[M=63, N=63]{\includegraphics[width=0.45\textwidth]{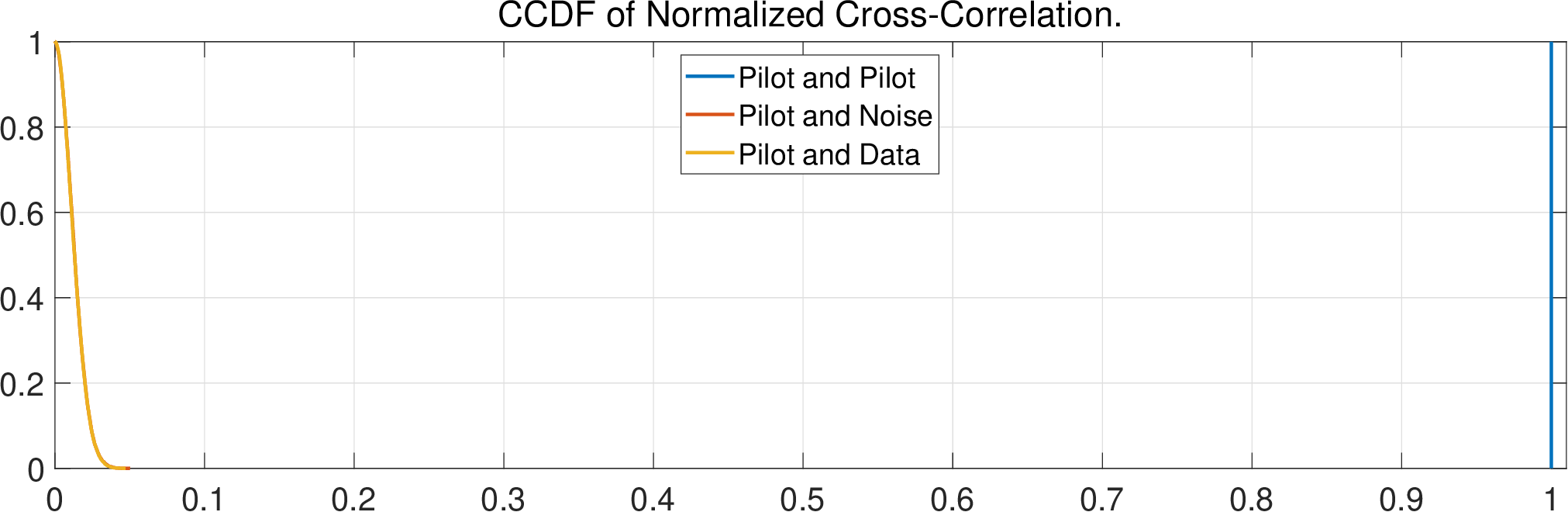}}\\
\subfigure[M=255, N=255]{\includegraphics[width=0.45\textwidth]{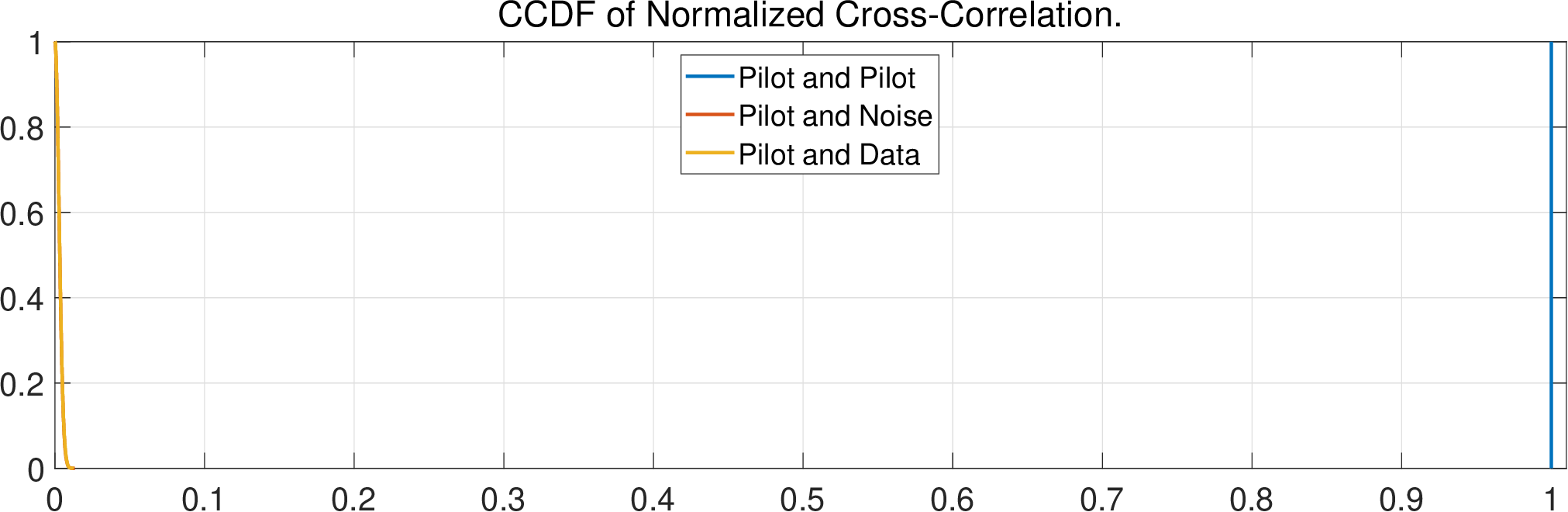}}\\
  \caption[]{Illustration of 2D pilot property.}
	\label{fig:fig1}
\end{figure}

Figure \ref{fig:fig1} shows the correlation property of the proposed 2D pilot. The blue curve with the magnitude value of one is the normalized auto correlation values of the proposed pilots with itself as a reference. The red and orange curves are the complementary cumulative distribution function (CCDF) of the normalized cross correlation values of the pilot with the QAM modulated data and noise samples, i.e., $\rho_{0i}$ and $\varsigma_{0i}$. It is observed that both the red and orange curves obey roughly the same bell-shaped distribution. Form figure \ref{fig:fig1}-(a) to \ref{fig:fig1}-(c), as the values of $M$ and $N$ increase from $15$ to $255$, the distribution of variables $\rho_{0i}$ and $\varsigma_{0i}$ concentrated towards smaller values, which is reflected in equation (\ref{eq:eq17}) as a better sensing SINR.

Based on above observation, we can claim that $Z$ is lower bounded and asymptotically increases to infinity when $MN$ goes to infinity. In other words, we can flexibly obtain different level of sensing service quality by simply scaling the dimension of the 2D pilot. 

%The system diagram of the proposed ISAC system is illustrated in figure \ref{fig:fig0}.
%\begin{figure}
%\centering
%\includegraphics[width=0.45\textwidth]{System_Diagram_w_Scrambler.eps}
  %\caption[]{Illustration of the system diagram.}
	%\label{fig:fig0}
%\end{figure}

\subsection{Fractional Doppler Refinement}
\begin{figure}
\centering
\includegraphics[width=0.40\textwidth]{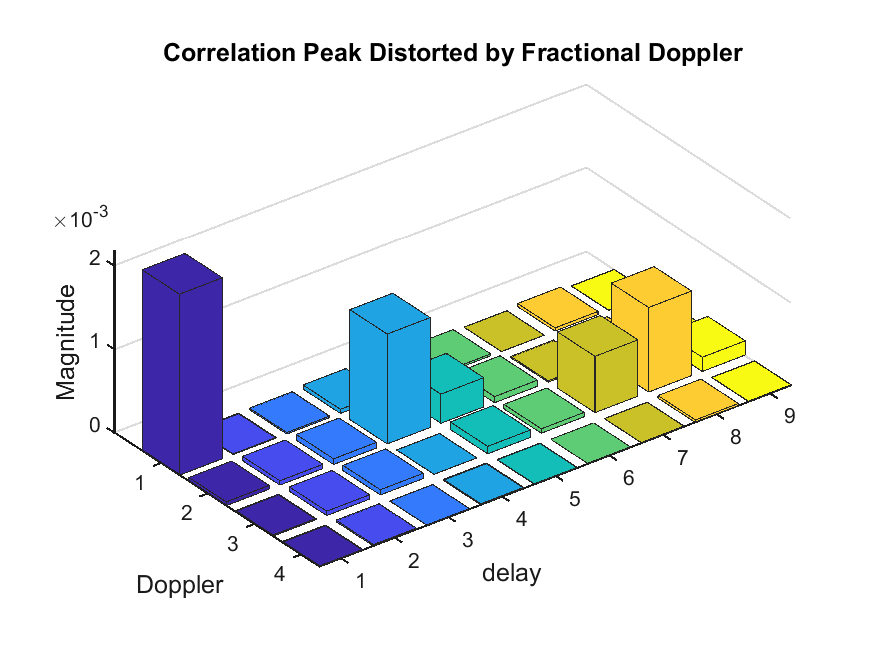}
  \caption[]{Impact of fractional Doppler.}
	\label{fig:fig0}
\end{figure}

Note that in algorithm \ref{alg:alg1} the detection of Doppler and delay is based on the shift-and-correlation operation. Therefore in theory only integer Doppler and delay can be detected. More specifically, the Doppler and delay granularity is limited to $k\frac{1}{NT}$ and $l\frac{1}{M\Delta f}$, which decreases with integer $M$ and $N$. Figure \ref{fig:fig0} illustrates the impact of fractional Doppler on algorithm \ref{alg:alg1}.

Suppose the three sensing targets in figure \ref{fig:fig0} are labeled with $1$, $2$, $3$ with the quantized Doppler and delay values of $<0,0>$, $<3.3,1>$ and $<6.5,2>$. It is noted that the correlation peak of target $1$ is not distorted and the power concentrates to $<0,0>$. The correlation peaks of target $2$ and $3$ are distorted and the power leaks to adjacent grids, and the level of leaked power proportional to the deviation from the integer Doppler. As a result, the power leakage achieves maximum at the fractional value $0.5$ and reduces to zero at $0$ or $1$. Based on these facts, we can decorate the integer output of the Doppler in algorithm \ref{alg:alg1} by exploiting the power of side peaks. Intuitively, a simple linear interpolation to refine the Doppler value can be summaries as,
\begin{align}
%\begin{split}
& \hat{\nu} = \sum^{J}_{j=1}{w_j\nu_{k_j}},
%\end{split}
\label{eq:eq17}
\end{align}
where $J$ adjacent peaks are chosen and $w_j$ is the weight factors proportional to the power $\bar{p}_j$ of peak $j$, $\nu_{k_j}$ is the integer Doppler corresponding to peak $j$. In this paper we heuristically choose the weight as $w_j = e^{\bar{p}_j}$. Note that the same operation can also be applied to fractional delay.

\section{Numerical Results}
In this section we provide the performance evaluation of the proposed sensing pilot underlaid ISAC system from the perspective of both the sensing and communication. We assume the sensing and communication receivers are not necessarily to be co-located such that the sensing and communication channels are irrelevant. 

For sensing detection we assume a multi-targets scenario where the three reflectors corresponds to three echo paths. For each echo path, different maximum Doppler $\tilde{\nu}_{i}$ is considered to generate the Doppler shift followed the uniform distribution $U(0, \tilde{\nu}_{i})$. For communication evaluation we directly adopt the power delay profile (PDP) of 3GPP extended vehicular A (EVA) models in \cite{3gppts36101} for demonstration of the symbol detection performance. 
The detailed system setup is shown in table \ref{tab:tab0}.
\begin{table}
\begin{center}
\caption{System Parameters}
\label{tab:tab0}
\resizebox{\linewidth}{!}{
\begin{tabular}{|c||c||c|}
\hline
    Parameter & Communication & Sensing\\
\hline
    Carrier frequency & 6GHz & 6GHz\\ 
\hline
    SCS & 60e3 & 60e3\\ 
\hline
    [$N$,$M$] & [16,64] & up to [512,64]\\ 
\hline
    Power Scale & 1 & 0.2\\
\hline
    PDP &  3GPP EVA &  [0, 0, 0]\\ 
\hline
    Modulation & 16QAM & BPSK\\ 
\hline
    Codec & LDPC-0.33 & M-sequence\\ 
\hline
    Velocity & 30 & 500\\ 
\hline
%\label{tab:tab0}
\end{tabular}}
\end{center}
\end{table}

We use the M-sequences generated by the different primitive polynomials as the component sequences of the 2D pilot. We assume that the allocated power ratio to the superimposed pilot and data symbols is $1:5$ respectively. We investigate the performance of sensing via the Doppler estimation error, and the performance of communication via bit error probability (BER).

\begin{figure}
\centering
\subfigure[Perfect CSI]{\includegraphics[width=0.38\textwidth]{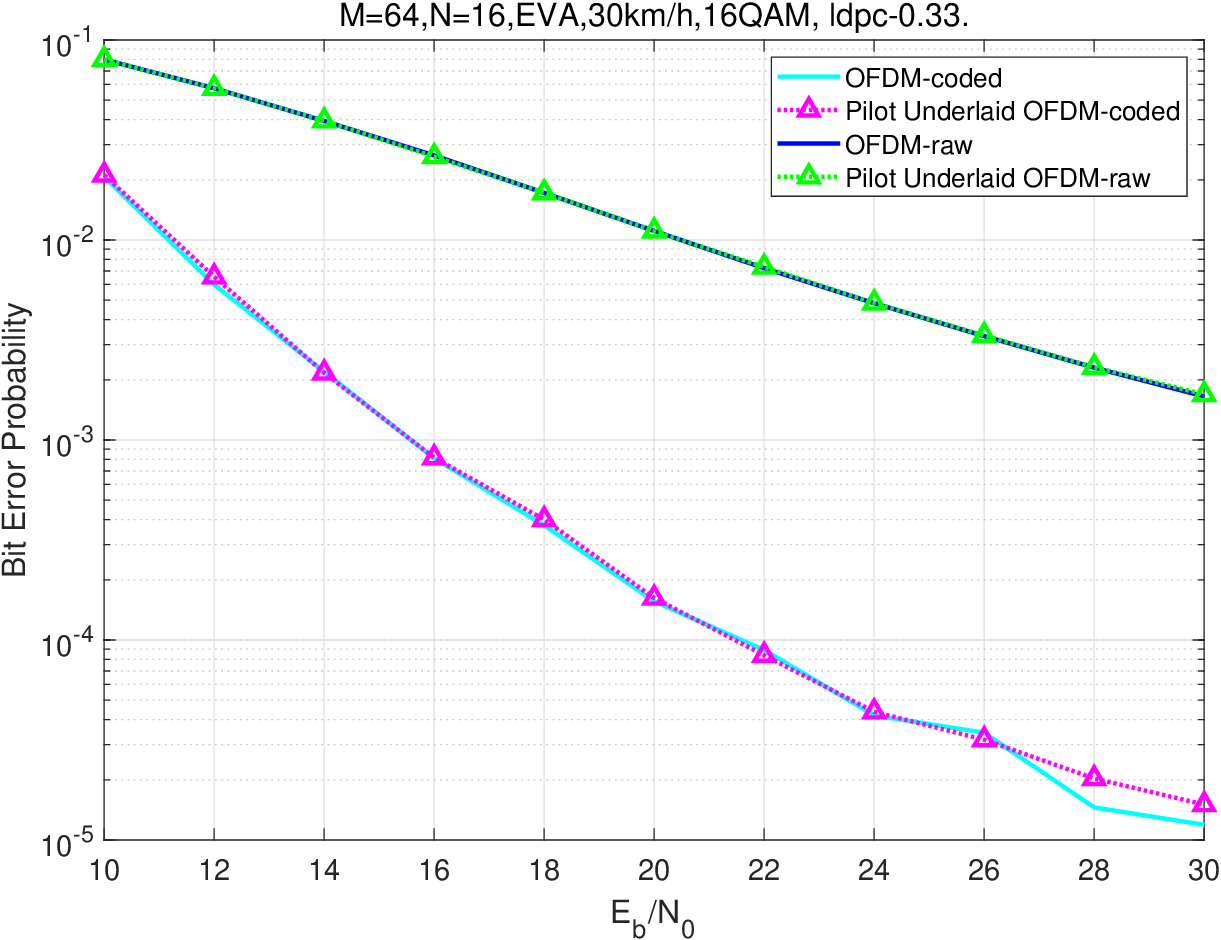}}\\
\subfigure[Estimated CSI]{\includegraphics[width=0.38\textwidth]{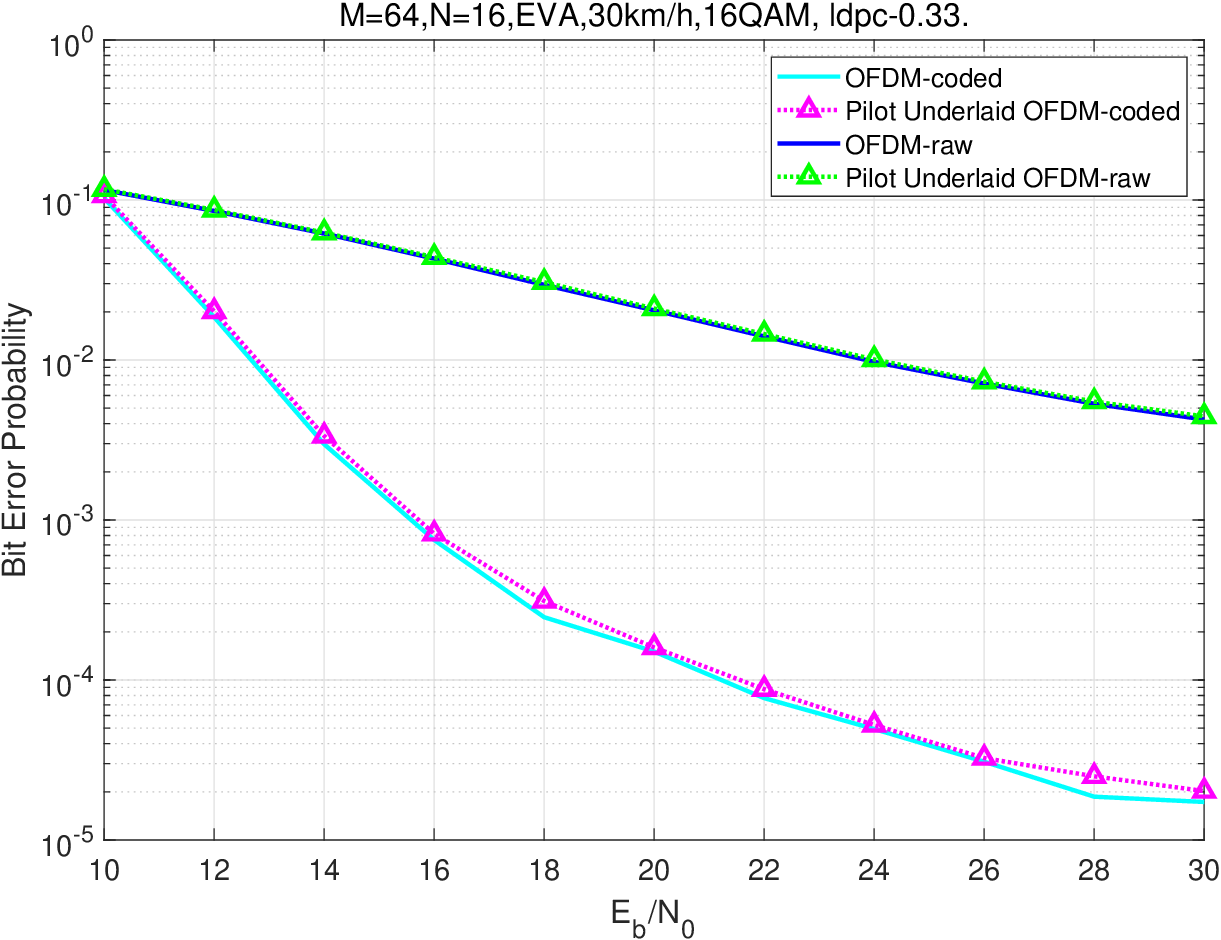}}\\
  \caption[]{Communication performance: bit error probability.}
	\label{fig:fig2}
\end{figure}

Figure \ref{fig:fig2} shows the comparison of the symbol detection performance of OFDM and sensing pilot underlaid OFDM (SPU-OFDM). The signal power of OFDM data is the same as the two alternatives while the power of additive sensing pilot is $0.2$ times of the data signal power. It is observed in figure \ref{fig:fig2}-(a) that the BER performance are almost identical except for high SNR region above $26$dB if perfect CSI is feasible. In figure \ref{fig:fig2}-(b), the BER performance of the sensing pilot underlaid OFDM is slightly worse as the channel estimation error will lead to larger residual interference in interference cancellation, which degrades the symbol detection performance. 

\begin{figure}
\centering
\includegraphics[width=0.38\textwidth]{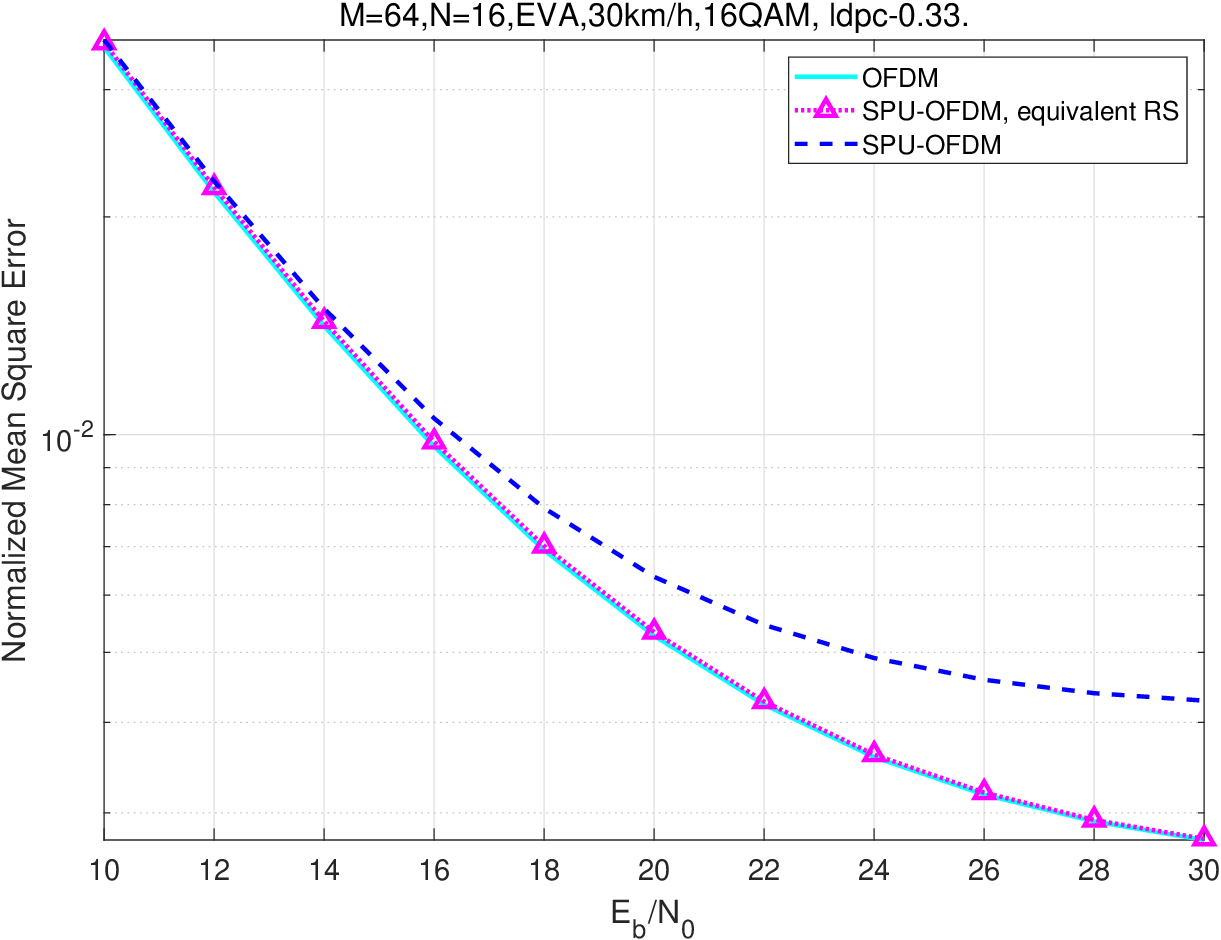}
%\subfigure[Estimated CSI]{\includegraphics[width=0.45\textwidth]{UnPL_EVA_30km_16qam_ecsi_nmse.eps}}\\
  \caption[]{Communication performance: channel estimation error.}
	\label{fig:fig3}
\end{figure}

Figure \ref{fig:fig3} depicts that the channel estimation accuracy is maintained as we use the equivalent RS constitute of both OFDM RS and the sensing pilot. The performance difference in terms of normalized mean square error (NMSE) between the OFDM and SPU-OFDM, i.e., the cyan solid line and magenta dotted curve with triangle, are negligible. In contrast, if we does not consider the contribution of the sensing pilot, then the blue dashed curve is much worse than the other two.

\begin{figure}
\centering
\includegraphics[width=0.42\textwidth]{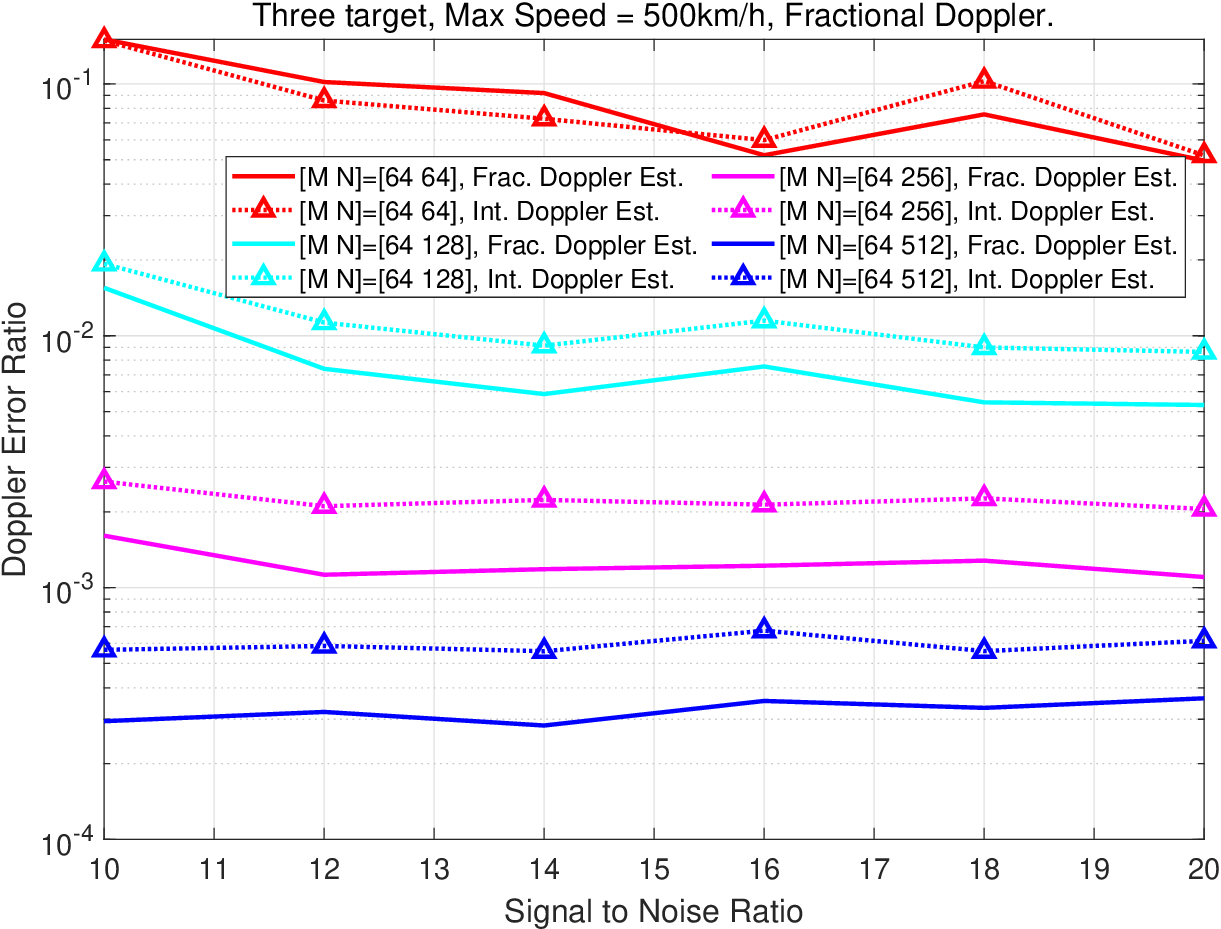}
  \caption[]{The comparison of sensing error in different pilot sizes.}
	\label{fig:fig4}
\end{figure}

Figure \ref{fig:fig4} demonstrates the performance of sensing detection against the dimension of the 2D pilot signal. For space limitation, we assume fractional Doppler only here and the fractional delay could be dealt in the same way. Here we define the Doppler error rate which is given by,
$
\frac{1}{p}\sum^p_{i=1}{|\frac{\hat{Dp}_{i}-Dp_{i}}{Dp_{i}}|},
$
where $Dp_{i}$ and $\hat{Dp}_{i}$ are the Doppler and its estimation of target $i$.  
It is observed that as $N$ increases from $64$ to $512$, the estimation error decreases dramatically, which coincides with our analysis in section \ref{sec:sec4}, i.e., increasing the scale of the pilot can suppress the distortion of the sensing detection. Furthermore, comparison results shows that the linear interpolation based on the peak power effectively improves the Doppler estimation accuracy. 

Another observation from figure \ref{fig:fig4} is that in the case of $N=256$ and $N=512$, the Doppler estimation error remains at the same level while in $N=128$ and $N=64$ the performance is visibly worse in low SNR region. The reason is that, when the scale of the sensing pilot is not large enough, the accumulated power of the correlation peak is not large enough to overwhelm the distortion brought by the noise and overlaid data. However, the impact of noise and data can be negligible when the pilot size is large enough. 

\section{Conclusion}
In this paper, we have presented and analyzed an underlaid 2D sensing pilot design for the ISAC system. We have demonstrated that the proposed design is compatible with concurrent wireless communication systems and offers flexibility in adapting to different sensing requirements. For sensing purposes, we have provided a low-complexity correlation-based sensing detection algorithm that leverages the twist-convolution property of the delay-Doppler domain channel. Furthermore, in the communication aspect, we have shown that the OFDM receiver can efficiently cancel the interference caused by the underlaid sensing pilot, leading to negligible negative impact on symbol detection. Our numerical results demonstrate that the underlaid pilot, along with an appropriate design and detection scheme, is a promising, efficient, and robust technique for multi-target detection in future ISAC systems.

\bibliographystyle{IEEEtran}
\bibliography{fmtc_TWC}
\end{document}